\documentstyle[twoside,epsf]{article} 
 
\input ibvs2.sty 
 
\begin{document} 
\IBVShead{5930}{12 March 2010} 
 
\IBVStitle{BVR$_{\rm C}$I$_{\rm C}$ photometric evolution and flickering during}
\IBVStitle{the 2010 outburst of the recurrent nova U Scorpii} 
  
\IBVSauth{U. Munari$^1$, S. Dallaporta$^2$, F. Castellani$^2$}

\IBVSinst{INAF Osservatorio Astronomico di Padova, Sede di Asiago, I-36032 Asiago (VI), Italy} 
\IBVSinst{ANS Collaboration, c/o Astronomical Observatory, 36012 Asiago (VI), Italy} 
 
\SIMBADobjAlias{V2673 Op}
\IBVStyp{Nova} 
\IBVSkey{photometry} 
\IBVSabs{CCD BVRcIc photometric observations of the 2010 outburst of the
recurrent nova U Scorpii are presented. The light-curve has a smooth
development characterized by t2(V)=1.8 and t3(V)=4.1 days, resembling that
of previous outbursts. The plateau phase in 2010 has been brighter, lasting
shorter and beginning earlier than in the 1999 outburst. Flickering, with an
amplitude twice large in $I_{\rm C}$  than in $B$ band, was absent on day +4.8
and +15.7, and present on day +11.8, with a time scale of about half an
hour.}

\begintext 
The 2010 outburst of the recurrent nova U Scorpii was discovered
by B.G. Harris (New Smyrna Beach, FL, USA) on Jan. 28.4385 UT,
when the star was measured at $V$=8.05 (cf Schaefer 2010). 
On Jan 27.63 UT, i.e. 0.80 days earlier, the nova was still
at quiescence brightness ($V$$\geq$16.5 mag, Linnolt 2010).

This is the 10th recorded outburst of U Scorpii. Previous ones
occurred on 1863, 1906, 1917, 1936, 1945, 1969, 1979, 1987 and 
1999 according to the recent summary by Schaefer (2009). The last
outburst has been the best observed one, with detailed reports 
being provided by Munari U. et al. (1999), Kiyota (1999), L{\'e}pine
et al. (1999), Anupama and Dewangan (2000), Hachisu et al. (2000), 
Evans et al. (2001) and Iijima (2002). 

We obtained accurate $B$$V$$R_{\rm C}$$I_{\rm C}$ of U Sco with a 0.30-m Meade
RCX-400 f/8 Schmidt-Cassegrain telescope equipped with a SBIG ST-9 CCD
camera.  The photometry was accurately corrected for color equations using
nightly calibration on Landolt (1992) standard stars.  The data are
presented in Table~1, and plotted in Figure~1.  The external errors (always
less than 0.02 mag) do not exceed the dimension of the symbols in Figure~1.

\IBVSfig{18cm}{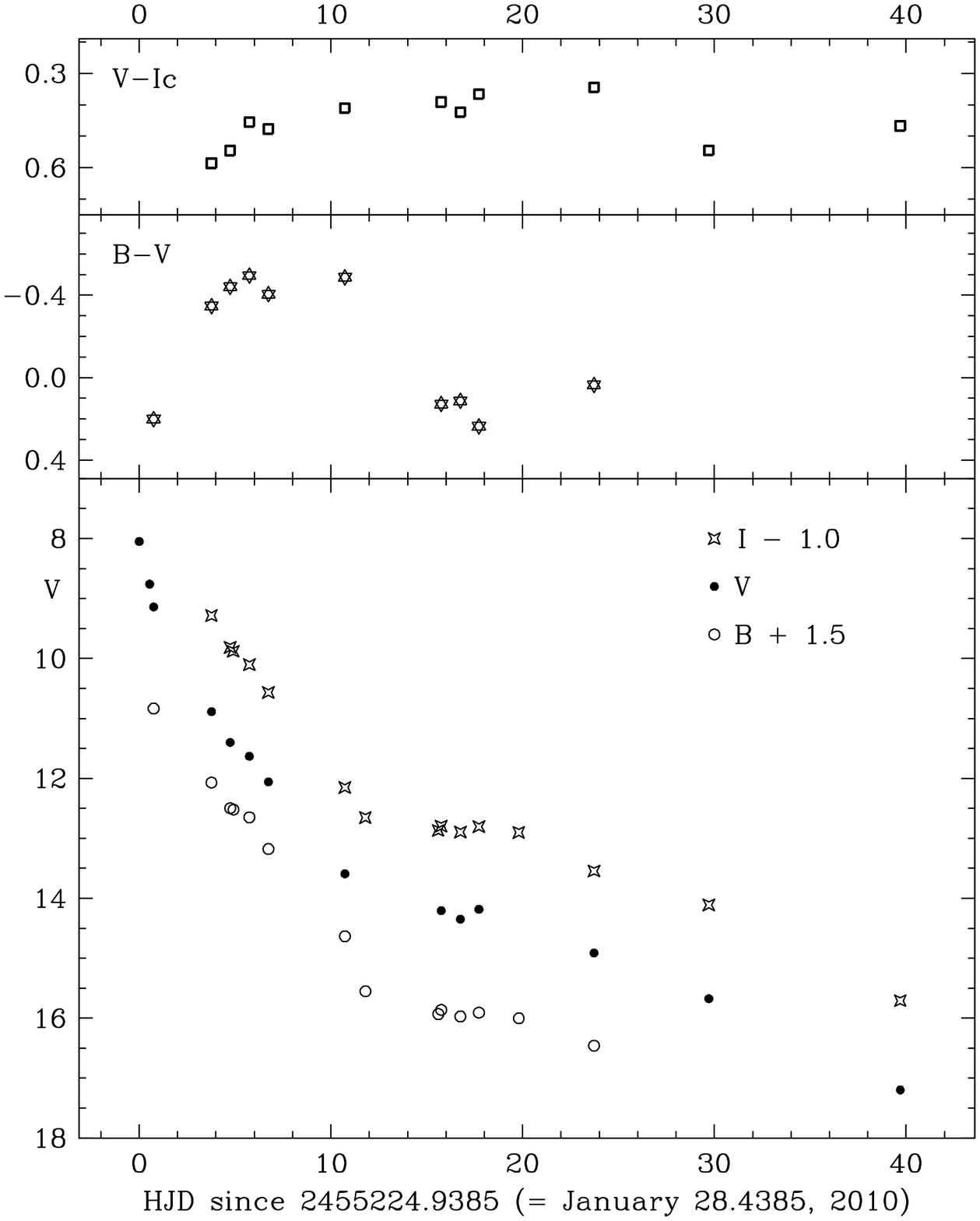}{$B$$V$$R_{\rm C}$$I_{\rm C}$ photometric
evolution of the 2010 outburst of the recurrent nova U Scorpii according
to our observations listed in Table~1. The point at $t$=0.0 days, $V$=8.05
is taken from IAUC~9111, that at $t$=0.54 days, $V$=8.76 from ATel 2412.}

In Figure~1 the time is counted from the discovery of U Sco in outburst
on Jan. 28.4385 UT ($t$=0.00 days), that we assume as the time of actual
maximum, there being no earlier observations of U Sco or reporting 
it brighter than $V$=8.05. The lightcurve in Figure~1 is charaterized by 
a smooth decline, similar to that of previous outbursts (cf Munari et
al. 1999; Kiyota 1999). The decline times ($\pm$0.1 days) are:
\begin{equation}
t_2^{V}=1.8   ~~~~~~~ t_3^{V}=4.1  ~{\rm days}
\end{equation}
that are significantly slower than $t_2$=1.2, $t_3$=2.6
days reported by Schaefer (2009) as typical values for previous outbursts,
and instead much closer to the $t_2$=2.2, $t_3$=4.3 days derived
by Munari et al. (1999) for the 1999 outburst. The light-curve in Figure~1
exhibits a plateau phase extending from day +12 to day +20, during which the
mean colors are
\begin{equation}
<V>=14.25 ~~<B-V>=+0.16 ~~<V-R_{\rm C}>=+0.35 ~~<V-I_{\rm C}>=+0.41
\end{equation}
This phase corresponds to the white dwarf still burning hydrogen in the
envelope and the ejecta being transparent to soft X-rays. In fact, on day
+12, Schlegel et al. (2010) found U Sco to have become a super-soft X-ray
source with a brightness 100 time larger than a previous observation on day
+8. Osborne et al. (2010) found U Sco still in super-soft conditions at day
+17.5.  A plateau was observed also during the 1999 outburst (Kiyota 1999, and
Hachisu et al. 2000), but at a fainter mean magnitude 
($<$$V$$>$=14.75), lasting slightly longer (11 days) and starting
appreciably later, on day +17.

With a 0.40-m f/8 Ritchey-Chr\'etien telescope located on Monte Baldo
(Verona, Italy), and equipped with a Finger Lake Instruments ML1001E CCD
camera, we carried out three runs in $B$ and $I_{\rm C}$ filters looking for
short time variations in U Sco.  The results are presented in Figure~2.  

\IBVSfig{16.5cm}{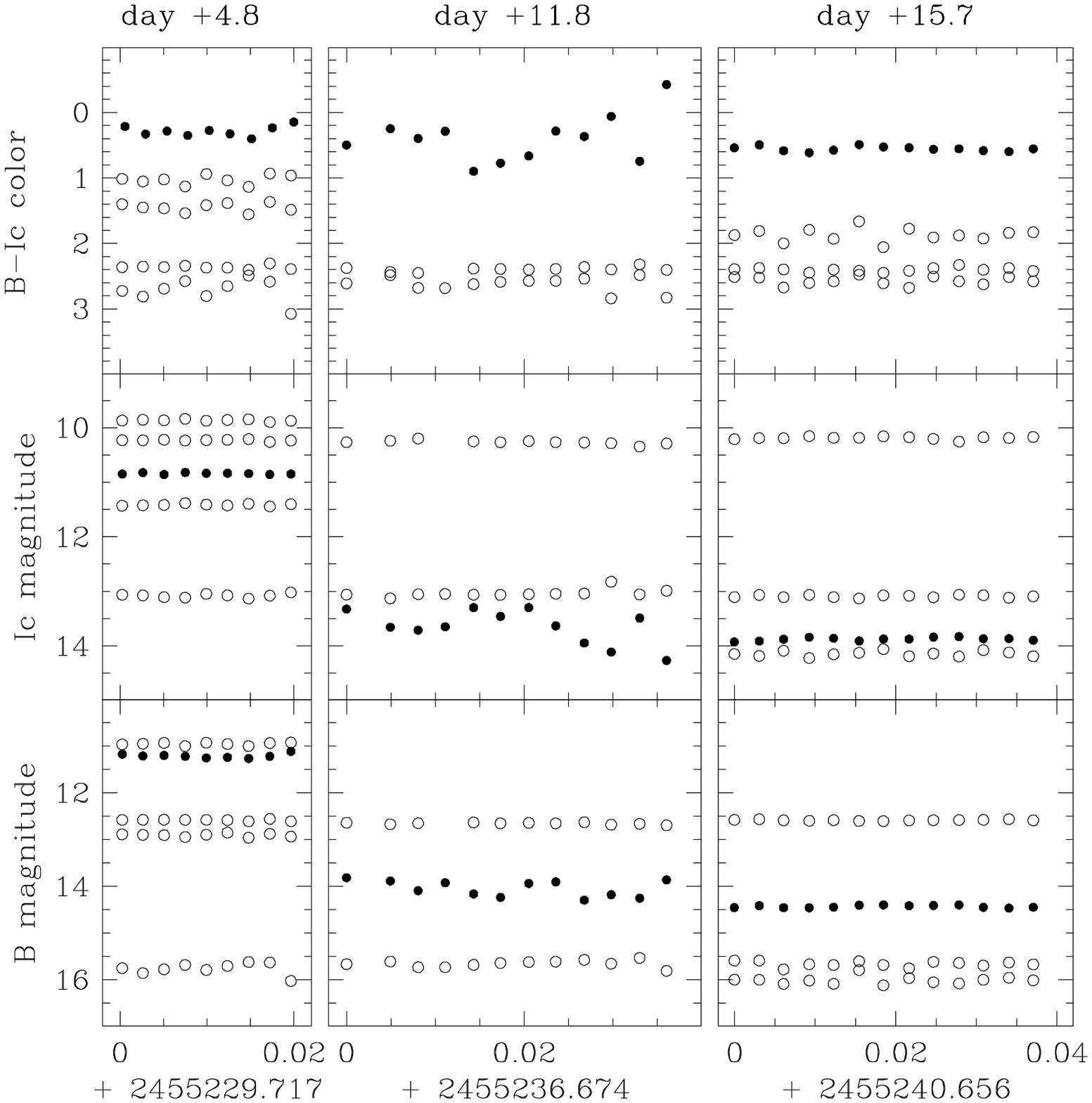}{Results of searches for flickering carried out on
days +4.8, +11.8 and +15.7 in the $B$ and $I_{\rm C}$ bands. The dots
represents measurements of U Sco, the circles those of nearby field stars
to monitor photometric stability.}

\vskip 3cm

\centerline{Table 1. Our $B$$V$$R_{\rm C}$$I_{\rm C}$ of U Scorpii}
\vskip 3mm
\begin{center}
\begin{tabular}{rrrrrrr}
\hline
\multicolumn{7}{c}{}\\
\multicolumn{1}{c}{HJD} &
\multicolumn{1}{c}{date HUT} &
\multicolumn{1}{c}{$V$} &
\multicolumn{1}{c}{$B$$-$$V$} &
\multicolumn{1}{c}{$V$$-$$R_{\rm C}$} &
\multicolumn{1}{c}{$V$$-$$I_{\rm C}$} &
\multicolumn{1}{c}{$R_{\rm C}$$-$$I_{\rm C}$}\\
\multicolumn{7}{c}{}\\
225.6933 &  2010 01 29.19  &  9.140  &   0.201  &         &         &           \\
228.7090 &  2010 02 01.21  & 10.888  & --0.347  &  0.831  &  0.586  & --0.300   \\
229.6861 &  2010 02 02.19  & 11.401  & --0.441  &  0.751  &  0.546  & --0.217   \\
230.6817 &  2010 02 03.18  & 11.632  & --0.495  &  0.582  &  0.455  & --0.239   \\
231.6799 &  2010 02 04.18  & 12.059  & --0.405  &  0.572  &  0.477  & --0.146   \\
235.6682 &  2010 02 08.17  & 13.592  & --0.487  &  0.365  &  0.410  &   0.022   \\
240.6921 &  2010 02 13.19  & 14.207  &   0.128  &  0.347  &  0.391  &   0.038   \\
241.6912 &  2010 02 14.19  & 14.349  &   0.113  &         &  0.423  &           \\
242.6575 &  2010 02 15.16  & 14.184  &   0.236  &  0.286  &  0.366  &   0.078   \\
248.6599 &  2010 02 21.16  & 14.912  &   0.035  &  0.275  &  0.344  &   0.062   \\
254.6492 &  2010 02 27.15  & 15.674  &          &  0.227  &  0.545  &   0.317   \\
264.6344 &  2010 03 09.13  & 17.197  &          &         &  0.467  &           \\
\multicolumn{7}{c}{}\\
\hline
\end{tabular}
\end{center}

\vskip 1cm

No
flickering was detected on day +4.8, while the short term variability was
clearly present on day +11.8 (at the beginning of the plateau phase). 
Worters et al.  (2010) reported that the flickering became visible on day
+8, and they attributed it to an accretion disk that had already been
re-established and was visible through optically thin ejecta.  Our last
observing run on day +15.7 (and additional three scattered points around day
+19.8), at the center of the plateau phase, did not however show any short
term variability, which cast doubts on the Worters et al.  interpretation in
terms of re-established accretion.  The flickering we observed had a
characteristic time scale of $\sim$half an hour, and a larger amplitude in
$I_{\rm C}$ ($\Delta m$=1.0 mag) than in $B$ band ($\Delta m$=0.5 mag).

\references 
 
 Anupama G.~C., Dewangan G.~C., 2000, AJ, 119, 1359 

 Evans A., Krautter J., Vanzi L., Starrfield S., 2001, A\&A, 378, 132 

 Iijima T., 2002, A\&A, 387, 1013 

 Hachisu I., Kato M., Kato T., Matsumoto K., 2000, ApJ, 528, L97 

 Kiyota S., 1999, IBVS, 4736, 1 

 Landolt, A.U. 1992, AJ 104, 340

 L{\'e}pine S., Shara M.~M., Livio M., Zurek D., 1999, ApJ, 522, L121 

 Linnolt, M. 2010, IAUC 9111

 Munari U., Zwitter, T., Tomov, T., et al., 1999, A\&A, 347, L39 

 Osborne, J.P., Page, K.L., Wynn, W. et al.  2010, ATel 2442

 Schaefer, B.E. 2009, arXiv:0912.4426

 Schaefer, B.E. 2010, IAUC 9111

 Schlegel, E.M., Schaefer, B.E., Pagnotta, A. et al. 2010, ATel 2430

 Worters, H.L., Eyres, S.P.S., Rushton, M.T. 2010, IAUC 9114

\endreferences

\end{document}